\documentclass[sigconf]{acmart}

\usepackage[table]{xcolor}
\usepackage{booktabs}
\usepackage{tabularx}
\usepackage{pifont} 

\definecolor{dmcallblue}{RGB}{235,243,250}

\AtBeginDocument{%
  }

\setcopyright{acmlicensed}
\copyrightyear{2018}
\acmYear{2018}
\acmDOI{XXXXXXX.XXXXXXX}

\acmConference[Conference acronym 'XX]{Make sure to enter the correct
  conference title from your rights confirmation email}{June 03--05,
  2018}{Woodstock, NY}

\acmISBN{978-1-4503-XXXX-X/2018/06}

\newcommand{\methodname}{\textbf{DM-Call}}

\begin{document}

\title{DM-Call: Decision-Model-Guided Function Calling with
Cumulative Verification Feedback}

\author{Minhyeong Yu}
\authornote{Both authors contributed equally to this research.}
\affiliation{%
  \institution{Enhans}
  \city{Seoul}
  \country{South Korea}}
\email{minhyeong@enhans.ai}

\author{Wonduk Seo}
\authornotemark[1]
\affiliation{%
  \institution{University College London}
  \city{London}
  \country{United Kingdom}}
\email{wonduk.seo.26@ucl.ac.uk}

\author{Seunghyun Lee}
\correspondingauthor
\affiliation{%
  \institution{Enhans}
  \city{Seoul}
  \country{South Korea}}
\email{seunghyun@enhans.ai}

\renewcommand{\shortauthors}{Minhyeong Yu\textsuperscript{*}, Wonduk Seo\textsuperscript{*}, Seunghyun Lee}

\begin{abstract}
Reliable function calling requires Large Language Models (LLMs) to select appropriate functions and generate arguments consistent with user intent. However, syntactically valid calls may still be semantically incorrect due to inappropriate function selection or mismatched arguments. In this paper, we propose \textbf{Decision-Model-Guided Function Calling (\methodname{})}, a framework that combines local schema validation with separate decision-model verification of function selection and argument correctness. For rejected candidates, \methodname{} accumulates verification probabilities, binary judgments, and schema errors across attempts. A hybrid feedback representation further combines numerical probabilities with verbal correctness labels to guide iterative refinement without executing external tools. Extensive evaluation across BFCL and API-Bank (first-call) demonstrates \methodname{} boosts function-calling accuracy versus direct generation, at modest additional cost.\footnote{Code is available at: github.com/gogoymh/decision-model-guided-function-calling}
\end{abstract}




\received{20 February 2007}
\received[revised]{12 March 2009}
\received[accepted]{5 June 2009}

\maketitle

\section{Introduction}
Function calling enables Large Language Models (LLMs) to interact with external tools and APIs by translating natural-language requests into structured function invocations~\cite{schick2023toolformer,li2023api}. Specifically, given a user request and available functions, an LLM determines whether a function call is needed and, if so, selects an appropriate function and generates arguments that reflect user intent. However, syntactically valid calls may still be semantically incorrect due to inappropriate function selection or mismatched arguments~\cite{patil2025berkeley}.

\begin{figure}[t]
    \centering
      \includegraphics[width=\linewidth]{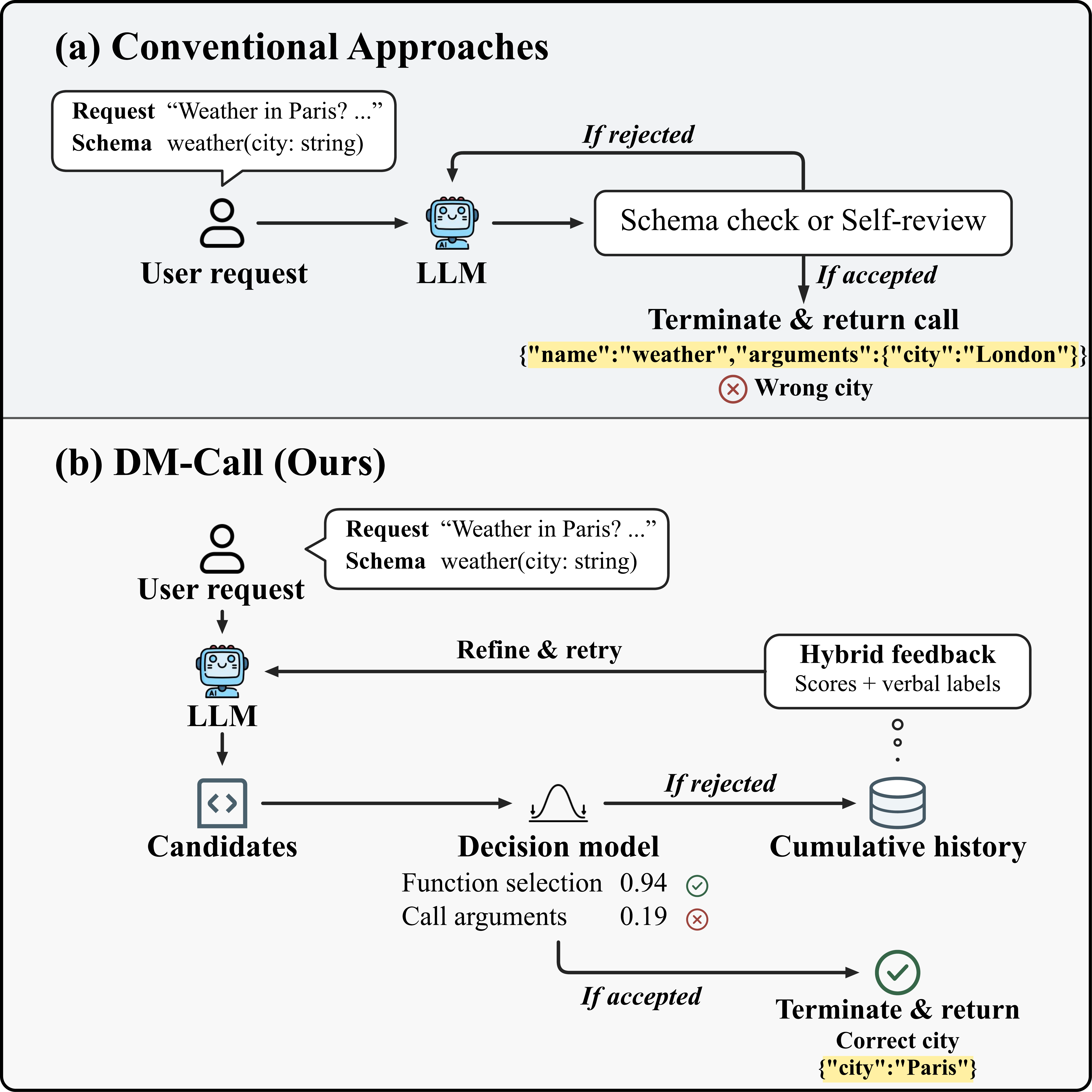}  
    \caption{
Comparison of conventional function-calling approaches and \methodname{}.
(a) Conventional approaches rely on schema validation or self-review, which may fail to detect semantically incorrect calls.
(b) \methodname{} separately verifies function selection and argument correctness through probabilistic judgments, accumulating rejected candidates and their hybrid feedback to guide subsequent revisions.
}
    \label{fig:DMCall_Square}
\end{figure}

Prior work has explored various approaches to improving function-calling reliability through validation and iterative refinement. Self-Refine revises candidate outputs using model-generated feedback~\cite{li2023api}. Other approaches, such as multi-agent debate and Reinforced Agent, incorporate feedback from multiple agents or a separate reviewer~\cite{du2023improving,ta2026reinforced}. More closely related, ToolVerifier applies self-verification to tool selection and parameter generation~\cite{mekala2024toolverifier}.

Despite this progress, several limitations remain: (1) syntactic constraints and schema validation alone cannot determine whether function selection and arguments align with user intent, (2) generic self-refinement and multi-agent approaches may incur additional inference overhead without explicitly distinguishing errors in function selection from those in argument generation, and (3) refinement strategies that rely solely on the latest feedback may overlook useful verification signals from earlier attempts.

To address these limitations, we propose \textbf{Decision-Model-Guided Function Calling (\methodname{})}, a framework that combines local schema validation with separate decision-model judgments of function selection and argument correctness, as illustrated in Figure~\ref{fig:DMCall_Square}. When a candidate fails verification, \methodname{} retains the rejected call along with its verification probabilities, binary judgments, and schema errors. To guide subsequent revisions, we introduce a \textit{hybrid feedback representation} that combines numerical verification scores with verbal correctness labels. By accumulating these signals across rejected attempts, \methodname{} enables the generator to iteratively refine candidate calls using feedback from previous failures.

We evaluate \methodname{} using four diverse LLMs and two decision-model verifiers on the Berkeley Function Calling Leaderboard (BFCL) single-turn cases~\cite{patil2025berkeley} and an adapted API-Bank first-call benchmark~\cite{li2023api}. \methodname{} improves function-calling accuracy and inference costs over baselines. Moreover, ablation studies demonstrate the benefits of cumulative feedback while examining verification components and feedback representations, alongside accuracy--latency trade-offs across repair budgets.

Our core contributions include (1) a decision-model-guided function calling framework that combines schema validation with separate verification of function selection and argument correctness, (2) a cumulative feedback mechanism that preserves rejected calls and their verification signals for iterative refinement, and (3) a hybrid verification-feedback representation that combines numerical verification probabilities with verbal correctness labels to guide subsequent revisions.

\begin{table*}[t]

\caption{Comparison of function-calling methods across four LLM generators on BFCL and API-Bank. We report three-run mean accuracy (\%) at $r^\star=3$ (s.d.\ 0.00--2.89) and total inference cost (USD). Decision. Generation refers to function-call generation guided by separate decision-model verification. Bold and underline indicate the best and second-best results per generator and metric, respectively; -- denotes unpriced local inference. $^{\dagger}$API-Bank uses first-call exact match without no-call negatives.}

\label{tab:main-table}
\centering\scriptsize
\setlength{\tabcolsep}{3pt}
\renewcommand{\arraystretch}{1.12}
\begin{tabular}{@{}l*{12}{c}@{}}
\toprule
& \multicolumn{3}{c}{\texttt{gpt-5-nano}}
& \multicolumn{3}{c}{\texttt{gpt-5.4-mini}}
& \multicolumn{3}{c}{\texttt{claude-sonnet-5.5}}
& \multicolumn{3}{c}{\texttt{gemma-4-31b-it}}\\
\cmidrule(lr){2-4}
\cmidrule(lr){5-7}
\cmidrule(lr){8-10}
\cmidrule(l){11-13}
Method
& BFCL$\uparrow$ & API-Bank$^{\dagger}\uparrow$ & USD$\downarrow$
& BFCL$\uparrow$ & API-Bank$^{\dagger}\uparrow$ & USD$\downarrow$
& BFCL$\uparrow$ & API-Bank$^{\dagger}\uparrow$ & USD$\downarrow$
& BFCL$\uparrow$ & API-Bank$^{\dagger}\uparrow$ & USD$\downarrow$\\
\midrule

\textbf{Direct Generation} & \multicolumn{12}{c@{}}{}\\
\addlinespace[2pt]
Direct
& $64.12$ & $61.32$ & $\mathbf{0.167}$
& $81.59$ & $\underline{80.13}$ & $\mathbf{2.146}$
& $88.25$ & $\underline{86.97}$ & $\mathbf{12.598}$
& $86.67$ & $81.41$ & --\\

Self-Refine
& $69.45$ & $58.55$ & $0.762$
& $81.27$ & $79.91$ & $9.591$
& $88.43$ & $\mathbf{87.39}$ & $56.184$
& $86.69$ & $81.41$ & --\\

\midrule
\textbf{Multi-Agent} & \multicolumn{12}{c@{}}{}\\
\addlinespace[2pt]
MAD-FC
& $63.53$ & $\underline{66.03}$ & $1.493$
& $81.10$ & $\mathbf{80.98}$ & $16.501$
& $85.22$ & $81.62$ & $95.244$
& $84.91$ & $\mathbf{82.48}$ & --\\

PF-FC
& $\underline{74.54}$ & $\mathbf{70.09}$ & $5.903$
& $81.79$ & $76.71$ & $6.750$
& $86.45$ & $83.12$ & $18.686$
& $84.85$ & $77.56$ & --\\

\midrule
\textbf{Decision. Generation} & \multicolumn{12}{c@{}}{}\\
\addlinespace[2pt]
\rowcolor{dmcallblue}
\methodname{} (\textit{Jev})
& $\mathbf{74.63}$ & $64.32$ & $\underline{0.525}$
& $\mathbf{82.54}$ & $\underline{80.13}$ & $\underline{2.783}$
& $\mathbf{88.56}$ & $\mathbf{87.39}$ & $\underline{13.988}$
& $\mathbf{86.91}$ & $\underline{81.62}$ & --\\

\rowcolor{dmcallblue}
\methodname{} (\textit{PPLX})
& $74.24$ & $64.53$ & --
& $\underline{82.20}$ & $\underline{80.13}$ & --
& $\underline{88.46}$ & $\mathbf{87.39}$ & --
& $\underline{86.88}$ & $81.41$ & --\\

\bottomrule
\end{tabular}
\end{table*}

\section{Methodology}
\subsection{Task Definition}
Given a user request $x$ and the available functions $\mathcal{F}$ with their schemas, a system returns one next action: a call $c=(f,a)$ with function $f\in\mathcal{F}$ and arguments $a$, or no call, $c=\varnothing$. The request may include earlier dialogue turns. The action is correct when it matches the benchmark reference, that is, the expected function with acceptable argument values, or no call when none is expected. 

It can be wrong in three specific ways: (1) it calls a function when no call is expected, or returns no call when one is; (2) it picks the wrong function; or (3) its arguments are wrong or missing. Because even a schema-valid wrong call can trigger an unintended external action, \methodname{} verifies and refines candidate calls before execution, without tool-execution feedback.

\subsection{Decision-Model Verification}

\methodname{} separates an LLM generator $G$, which proposes candidate calls, from a decision model $D$, which verifies them. At iteration $t$, the generator produces $c_t$ conditioned on the request, function schemas, and a history $H_t$ of rejected candidates and their feedback. The decision model $D$ asks two yes/no questions about $c_t$ and returns the probability of \emph{yes} for each, $p_f^{(t)}, p_a^{(t)} \in [0,1]$:
\begin{equation}
c_t=G(H_t,x,\mathcal{F}),\qquad
p_f^{(t)}=D_f(x,\mathcal{F},c_t),\quad
p_a^{(t)}=D_a(x,\mathcal{F},c_t).
\label{eq:generate}
\end{equation}
The function question $D_f$ assesses whether the selected function matches the request or, for a no-call candidate proposed during repair, whether no call is appropriate. The argument question $D_a$ assesses whether the required values are present, supported by the request, and assigned to the correct parameters. For a no-call candidate, argument verification is skipped and $p_a^{(t)}=\bot$ (a null value). A parseable call with missing arguments still receives both judgments. As an exception, if the initial candidate $c_0$ is a parseable no-call, \methodname{} returns it without verification or repair and marks it as unverified. An unparseable output is not treated as a no-call.

\subsection{Acceptance and Cumulative Repair}
\label{sec:acceptance}
\methodname{} accepts a candidate only when it passes a schema check and every decision-model question that applies to it. The schema check $S(c_t)$ is true when $c_t$ is readable, names a function in $\mathcal{F}$ and supplies every required argument with its declared type. A question that is not asked, such as the argument question on a no call or any question on an unreadable candidate, has the missing score $\bot$. The verdict $V(p)$ is true if and only if $p\neq\bot$ and $p\ge0.5$. Candidate $c_t$ is accepted, written $A_t=\text{true}$, when
\begin{equation}
A_t=\begin{cases}
S(c_t)\land V\bigl(p_f^{(t)}\bigr), & c_t\ \text{is a no call},\\
S(c_t)\land V\bigl(p_f^{(t)}\bigr)\land V\bigl(p_a^{(t)}\bigr), & \text{otherwise},
\end{cases}
\label{eq:gate}
\end{equation}
where $\land$ denotes logical AND. A readable candidate that fails $S$ is still judged by $D$, so its feedback carries both the schema errors and the decision model's judgments.

When $c_t$ is rejected ($A_t$ is false) and a repair remains ($t<r$), \methodname{} queries the generator for a revised candidate $c_{t+1}=G(H_{t+1},x,\mathcal{F})$ (Eq.~\ref{eq:generate}). The history $H_t$ lists the candidates rejected before step $t$, oldest first, each paired with its feedback $d_i$:
\begin{equation}
\begin{aligned}
H_t&=\bigl[(c_0,d_0),\dots,(c_{t-1},d_{t-1})\bigr],\\
d_i&=\bigl(\Phi(p_f^{(i)}),\,\Phi(p_a^{(i)}),\,V(p_f^{(i)}),\,V(p_a^{(i)}),\,S(c_i)\bigr).
\end{aligned}
\label{eq:history}
\end{equation}

Here $\Phi$ denotes the feedback representation used to convey verification scores to the generator, while the remaining entries capture the binary judgments and schema validation results, including any error messages.\footnote{Section~\ref{sec:hybrid} describes the default \emph{hybrid} feedback representation.} The generator receives these records, explicitly marked as unexecuted, alongside the original request. In contrast, the decision model evaluates only the current candidate against the request and function schemas. \methodname{} returns the first accepted candidate or falls back to the initial candidate $c_0$ if none passes verification within the repair budget.

\subsection{Hybrid Feedback}
\label{sec:hybrid}

The feedback representation $\Phi$ in Eq.~\ref{eq:history} determines how verification scores are conveyed to the generator. Our default representation, \emph{Hybrid}, pairs each probability with a verbal label:
\begin{equation}
\begin{aligned}
\Phi(p) &= \bigl(p,\mathcal{L}_{k(p)}\bigr), \quad
\Phi(\bot) = (\bot,\bot),\\
k(p) &= \min\bigl(\lfloor 5p \rfloor + 1,\,5\bigr).
\end{aligned}
\label{eq:hybrid-feedback}
\end{equation}

Here, $\mathcal{L}=(L_1,\ldots,L_5)$ denotes the ordered labels \emph{very likely incorrect}, \emph{likely incorrect}, \emph{uncertain}, \emph{likely correct}, and \emph{very likely correct}. Each label corresponds to a probability bin of width $0.2$, ordered by increasing $p$, with $p=1$ assigned to the final bin. The representation affects only the feedback provided to the generator, while verification decisions continue to use the numeric probabilities through $V$ (Eq.~\ref{eq:gate}).\footnote{Section~\ref{sec:feedback-representation} compares \emph{Hybrid} with \emph{Numeric} and \emph{Text} feedback.}

\section{Experiments}
\subsection{Experimental Setup}
\label{sec:data}

\paragraph{Datasets and Metrics.}
For BFCL, we use 3,035 single-turn Python cases, including irrelevance cases that expect no call, scored by the official AST metric.\footnote{We exclude 16 live-relevance cases that test only whether a call is made.} For API-Bank~\cite{li2023api}, we use 156 first-call test cases with a fixed 53-API catalog, scored by exact match of function names and type-normalized arguments without executing APIs.

\paragraph{Models.}

We evaluate four LLM generators: (i) \texttt{gpt-5-nano}, (ii) \texttt{gpt-5.4-mini}~\cite{singh2025openai}, (iii) \texttt{claude-sonnet-5.5}\footnote{\url{https://www.anthropic.com/claude-sonnet-5-5}}, and (iv) the 31B instruction tuned variant of \texttt{gemma-4}~\cite{team2026gemma}. 
Reasoning effort is set to \texttt{none} where supported, with no additional thinking mode enabled. For verification, we employ two decision models: the hosted \textit{Jev} (\texttt{jev-1.13.0})\footnote{https://typesafe.ai/blog/introducing-system-one-models-and-jev} and the open-weight 27B \textit{PPLX Decider v1.1}~\cite{nimble2026}.

\paragraph{Baselines and DM-call (Ours)}
We consider two baseline categories (Table~\ref{tab:main-table}).
\emph{Direct Generation} includes (i) \emph{Direct}, which returns the initial prediction, and (ii) \emph{Self-Refine}~\cite{madaan2023self}, which iteratively revises predictions using self-generated critiques for up to three rounds.
\emph{Multi-Agent} includes (iii) \emph{MAD-FC}~\cite{du2023improving}, which uses three agents for one debate round with majority voting, and (iv) \emph{PF-FC}~\cite{ta2026reinforced}, which employs an LLM reviewer for up to three refinement rounds.\footnote{All methods except MAD-FC share the initial Direct prediction; MAD-FC also recognizes tool calls expressed as text.}

For \methodname{}, we employ separate function and argument verification with a threshold of $0.5$ and up to three repair rounds ($r^\star=3$).

\subsection{Main Results}

\methodname{} achieves the highest BFCL accuracy across all four generators (Table~\ref{tab:main-table}), with particularly substantial gains for \texttt{gpt-5-nano} (+10.5 percentage points over Direct). In contrast, conventional refinement and multi-agent baselines show less consistent improvements, often at higher inference costs. On API-Bank, performance gains are more limited, with \methodname{} achieving competitive results while several baselines perform better on individual generators. Overall, \methodname{} offers a favorable accuracy--cost trade-off, particularly for weaker generators, by selectively refining candidates that fail verification.

\subsection{Ablation Study}
\label{sec:ablation}

\begin{table}[t]
\caption{Ablation study of \methodname{} on BFCL using \texttt{gpt-5-nano} and \textit{Jev} ($r=3$), reporting three-run mean accuracy (\%) $\pm$ standard deviation for overall function calling, function selection, and argument correctness. Bold and underline indicate the best and second-best results per metric, respectively.}
\label{tab:ablation}
\centering\scriptsize
\setlength{\tabcolsep}{10pt}
\begin{tabular}{lrrr}
\toprule
Method & Call (\%) & Func. (\%) & Args. (\%)\\
\midrule
Direct & $64.11\pm0.31$ & $75.11\pm0.29$ & $81.27\pm0.14$\\
\textbf{(i) DM-Call-Function} & $73.12\pm0.46$ & $\underline{83.84}\pm0.52$ & $81.78\pm0.07$\\
\textbf{(iii) DM-Call-Argument} & $\underline{73.61}\pm0.33$ & $83.73\pm0.52$ & $\mathbf{82.64}\pm0.38$\\
\textbf{(iii) DM-Call-Latest} & $73.05\pm0.24$ & $83.61\pm0.22$ & $82.05\pm0.16$\\
\rowcolor{dmcallblue}
\textbf{DM-Call (Default)} & $\mathbf{74.32}\pm0.22$ & $\mathbf{84.80}\pm0.26$ & $\underline{82.18}\pm0.13$\\
\bottomrule
\end{tabular}
\end{table}

\begin{table}[t]
\caption{Comparison of Numeric, Text, and Hybrid feedback representations. Results show three-run mean accuracy (\%) (s.d.\ 0.02--0.54), with overall-averages across generators. Bold and underline indicate the highest and second-highest average scores per decision model, respectively.}
\label{tab:feedback-transfer}
\centering\scriptsize
\setlength{\tabcolsep}{6pt}
\begin{tabular}{lrrrrrr}
\toprule
& \multicolumn{3}{c}{\textit{Jev}}
& \multicolumn{3}{c}{\textit{PPLX}}\\
\cmidrule(lr){2-4}
\cmidrule(l){5-7}
Generator & Numeric & Text & Hybrid & Numeric & Text & Hybrid\\
\midrule
\texttt{gpt-5-nano} & $74.20$ & $74.14$ & $74.63$ & $73.95$ & $73.68$ & $74.24$\\
\texttt{gpt-5.4-mini} & $82.45$ & $82.47$ & $82.54$ & $82.22$ & $82.26$ & $82.20$\\
\texttt{claude-sonnet-5.5} & $88.53$ & $88.57$ & $88.56$ & $88.45$ & $88.46$ & $88.46$\\
\texttt{gemma-4-31b-it} & $86.93$ & $86.91$ & $86.91$ & $86.89$ & $86.91$ & $86.88$\\
\midrule
\rowcolor{dmcallblue}
\textbf{Avg. Score} & $\underline{83.03}$ & $83.02$ & $\mathbf{83.16}$ & $\underline{82.88}$ & $82.83$ & $\mathbf{82.95}$\\
\bottomrule
\end{tabular}
\end{table}

\paragraph{Verification and history.}
Table~\ref{tab:ablation} examines the contributions of separate verification and cumulative feedback. \textbf{(i) DM-Call-Function} and \textbf{(ii) DM-Call-Argument} evaluate only function selection or argument correctness, respectively, while \textbf{(iii) DM-Call-Latest} retains only the most recent rejection. All variants outperform Direct by $8.9$--$10.2$ percentage points. The full \methodname{} achieves the highest overall call accuracy with low standard deviations across runs, whereas \textbf{(ii) DM-Call-Argument} yields the best argument accuracy. Using only the latest feedback reduces call accuracy by $1.3$ points, highlighting the benefit of retaining earlier verification feedback.

\paragraph{Feedback representation.}
\label{sec:feedback-representation}
Table~\ref{tab:feedback-transfer} compares three feedback representations: \emph{Numeric}, which provides verification probabilities; \emph{Text}, which uses the verbal labels in Eq.~\ref{eq:hybrid-feedback}; and \emph{Hybrid}, which combines both. Hybrid yields the largest gains for \texttt{gpt-5-nano}, where repairs are more frequent ($24\%$ with \textit{Jev} and $21\%$ with \textit{PPLX}), outperforming Numeric by $0.3$--$0.4$ points and Text by $0.5$--$0.6$ points. For the remaining generators, which require fewer repairs ($1$--$6\%$), differences among the representations remain within $0.1$ points. Overall, Hybrid achieves the highest average accuracy with both decision models, supporting its use as the default representation.

\paragraph{Repair budget.}
\label{sec:repair-budget}
\begin{figure}[t]
    \centering
    \includegraphics[width=1\linewidth]{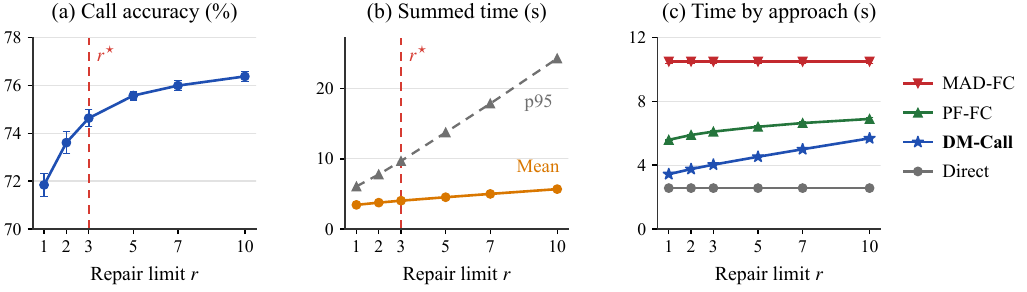}
    \caption{Effect of repair budget $r$ on \methodname{} using \texttt{gpt-5-nano}, BFCL, and \textit{Jev}. (a) Call accuracy, (b) mean and p95 latency, and (c) latency comparison with baselines.}
    \label{fig:repair}
\end{figure}

Figure~\ref{fig:repair} shows the accuracy--latency trade-off across repair budgets. Increasing $r$ from $3$ to $10$ improves accuracy by only $1.8$ points, while p95 latency rises from $9.7$ to $24.3$\,s. At $r=3$, \methodname{} achieves a mean latency of $4.0$\,s, compared with $2.6$\,s for Direct, $6.1$\,s for PF-FC, and $10.5$\,s for MAD-FC. We select $r^\star$ using an elbow-style criterion, $r^\star=\arg\max_{r\in R}(\widetilde{\mathrm{Acc}}_r-\widetilde C_r)$, where $R=\{1,2,3,5,7,10\}$ and $\widetilde X_r=(X_r-X_1)/(X_{10}-X_1)$ normalizes accuracy and requests per case. This yields $r^\star=3$; increasing to $r=5$ adds only $0.9$ points but requires $16\%$ more requests and $42\%$ higher p95 latency.

\section{Case Study}
\label{sec:case-study}

\begin{table}[!tb]
\caption{Case studies of DM-Call with \textit{Jev} at $r^\star=3$, illustrating function correction, argument refinement, no-call decisions, and multi-step repair. Each step reports prob. ($p_f$, $p_a$) and correctness, with -- indicating skipped argument verification.}
\label{tab:cases}
\centering\scriptsize
\setlength{\tabcolsep}{3pt}
\newcommand{\casehead}[1]{\multicolumn{5}{@{}>{\raggedright\arraybackslash}p{\columnwidth}@{}}{#1}}
\begin{tabularx}{\columnwidth}{@{}c>{\raggedright\arraybackslash}Xrrc@{}}
\toprule
$t$ & Proposed call $c_t$ & $p_f$ & $p_a$ & Correct\\
\midrule
\casehead{\textit{(a) Function.} \texttt{claude-sonnet-5.5}, BFCL: ``how to cook sikhae''}\\
0 & \texttt{HNA\_WQA\_search(keyword="how to make sikhae Korean sweet rice drink recipe")} & $0.37$ & $0.94$ & \ding{55}\\
1 & \texttt{cookbook\_search\_recipe(keyword="sikhae")} & $0.93$ & $0.95$ & \ding{51}\\
\midrule
\casehead{\textit{(b) Argument value.} \texttt{gpt-5.4-mini}, BFCL: ``Find me the best Italian restaurants in New York City with average customer ratings of more than 4 and accepts credit cards.''}\\
0 & \texttt{restaurant\_search(accepts\_credit\_cards=true, cuisine="Italian", location="New York City, NY", rating=5)} & $0.94$ & $0.19$ & \ding{55}\\
1 & \texttt{restaurant\_search(\ldots, rating=4)} & $0.96$ & $0.57$ & \ding{51}\\

\midrule
\casehead{\textit{(c) No call.} \texttt{gpt-5-nano}, BFCL: a question in Chinese, ``Who is Musk?'', with only a tax calculator}\\
0 & \texttt{calculate\_tax(filing\_status="single", income=0)} & $0.03$ & $0.12$ & \ding{55}\\
1 & no call; answers in text & $0.97$ & -- & \ding{51}\\
\midrule
\casehead{\textit{(d) Several repairs.} \texttt{gpt-5-nano}, API-Bank: ``Can you help me delete my account?'' \ldots\ ``My username is `foo', password is `bar'.'' Assistant: ``Alright, let me check if the token you provided is valid.''}\\
0 & \texttt{CheckToken(token="")} & $0.28$ & $0.07$ & \ding{55}\\
1 & \texttt{CheckToken(token="")} & $0.33$ & $0.06$ & \ding{55}\\
2 & \texttt{DeleteAccount(token="")} & $0.43$ & $0.06$ & \ding{55}\\
3 & \texttt{GetUserToken(username="foo", password="bar")} & $0.84$ & $0.96$ & \ding{51}\\

\bottomrule
\end{tabularx}
\end{table}

Table~\ref{tab:cases} presents four representative repair trajectories of \methodname{} with \textit{Jev}. The cases illustrate how verification guides refinement. In (a), a low $p_f$ redirects Sonnet to the correction, while in (b), a low $p_a$ prompts \texttt{gpt-5.4-mini} to correct the rating threshold. In (c), \texttt{gpt-5-nano} switches to a no-call response, whereas (d) demonstrates multi-step refinement leading to the correct call.

\section{Conclusion}
\methodname{} combines separate decision-model verification of function selection and argument correctness with iterative refinement using cumulative Hybrid feedback, improving both function-calling accuracy and cost-effectiveness.

\bibliographystyle{ACM-Reference-Format}
\bibliography{sample-base}

\end{document}